\documentclass{calor2002}

\begin{document}

\title{Setting the jet energy scale for the ZEUS calorimeter}

\author{M. Wing (On behalf of the ZEUS collaboration)}

\address{Bristol University, DESY F1, Notkestrasse 85\\
22607 Hamburg, GERMANY\\
E-mail: wing@mail.desy.de}

\maketitle

\abstracts{
A much improved determination of the transverse energy of jets has been
carried out in ZEUS, using a correction procedure based on two independent
methods. The first is based on a combination of tracking and
calorimeter information which optimises the resolution of reconstructed
kinematic variables. The conservation of energy and momentum in neutral
current deep inelastic $e^+p$ scattering events is exploited to determine
the energy corrections by balancing the kinematic quantities of
the scattered positron with those of the hadronic final state. The method
has been independently applied to data and simulated events. The second
method uses calorimeter cells as inputs to the jet
algorithm. Simulated events are then used to provide a correction for
the energy loss due to inactive material in front of the calorimeter.
A detailed comparison of the jet transverse energy and the transverse
energy of tracks in a cone around the jet provides the final correction.
This procedure relies on an accurate simulation of charged tracks and so 
is less reliant on simulating the energy loss of neutral particles in
inactive material. Final comparisons of the data and simulated events for
both methods allow an uncertainty $\pm 1\%$ to be assigned to the jet energy
scale.}

\section{Introduction}

The energy scale uncertainty of the calorimeter (CAL) coupled with differences between 
data and Monte Carlo (MC) simulations has traditionally been the dominant systematic 
uncertainty in jet measurements from the ZEUS collaboration. Energy scale uncertainties 
of $\pm (3-5)\%$, lead to uncertainties of $\sim \pm (10-20)\%$ in the cross-section 
measurements\cite{old_jet_papers}. 

The HERA accelerator collides positrons of 27.5~GeV with protons of \mbox{820 (or 920) GeV} 
leading to heavily boosted final states. Neutral current deep inelastic scattering events 
with high momentum transfer, $Q^2$, provide both interesting physics and the opportunity 
to study and calibrate the CAL energy scale. Quark-parton model type events, in which the 
positron scatters off a quark in the proton producing a final state jet back-to-back with 
the scattered positron, have been selected. After setting the electromagnetic energy scale, 
as discussed in section~\ref{sec:em}, a comparison of the scattered positron energy with 
that of the hadronic jet allows the determination of the uncertainty on the hadronic energy 
scale as discussed in the rest of this paper.

\section{Electromagnetic energy scale uncertainty}
\label{sec:em}

\begin{figure}[th]
\centerline{\epsfxsize=7cm\epsfbox{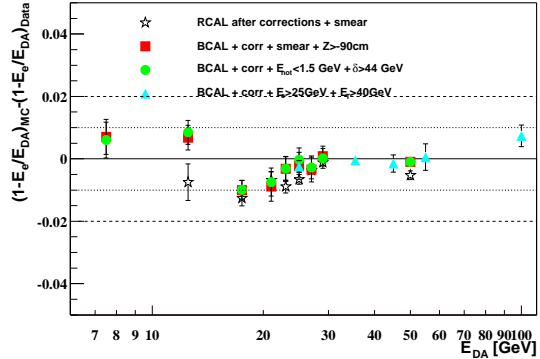}}
\caption{ Difference between data and MC of the energy scale for scattered positrons as 
a function of the energy, $E_{\rm DA}$. \label{fig:electron}}
\end{figure}

The electromagnetic energy scale has been studied in detail\cite{desy2001-055} 
by taking the ratio of the energy of the scattered positron measured in the calorimeter, 
$E_e$, with the track momentum or the electron energy reconstructed via the double angle (DA) 
method, $E_{\rm DA}$. The DA method predicts the electron energy from the angular 
information of the scattered positron and the hadrons and is, therefore, to first order, 
independent of the absolute energy scale of the CAL\cite{DA}. The difference between 
the energy ratio in data and MC is shown in figure~\ref{fig:electron}. The agreement is 
within $\sim \pm 1\%$. 

\section{Jet energy scale uncertainty}

Due to inactive material between the interaction point and the CAL, jets need to be 
corrected for the energy loss. Typically, $20\%$ of the jet's transverse energy 
is lost and is the major factor to be accounted for in order to produce an accurate 
determination of the uncertainty in the jet energy scale. Two different methods for 
correction have been developed.

\subsection{Method 1}

The first method uses energy flow objects (EFOs) to reconstruct the final state in which a 
combination of tracking and calorimeter information is used\cite{briskin}. Clusters 
of cells are formed and combined with tracks originating from the primary vertex and a 
decision made on whether to use the cluster or track. In the case of isolated clusters 
or tracks, the decision is trivial. For a matched cluster-track system, the resolutions 
of each object and ratio of energy to momentum are considered. Using this procedure, a 
list of track-EFOs and CAL-EFOs was obtained, where the track-EFOs are assumed to be an 
accurate measure of the particle energy and the CAL-EFOs are subject to energy loss in the 
inactive material and must, therefore, be corrected.

\begin{figure}[thp]
\centerline{\epsfxsize=8cm\epsfbox{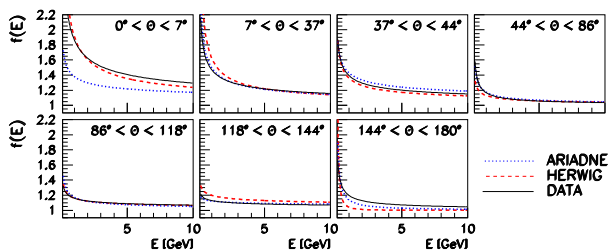}}
\caption{Energy corrections as a function of cluster energy in bins of $\theta$. The 
corrections are shown separately for data (solid line) and {\sc Herwig} MC (dashed line) 
and {\sc Ariadne} MC simulations (dotted line). \label{fig:function}}
\end{figure}

The conservation of energy and momentum in NC events was exploited to determine the
CAL-EFO energy-correction functions by balancing the momentum of the scattered positron with
that of the hadronic final state\cite{joost,andreas,desy01220}. Two samples of events were 
used, both with $Q^2>100$ GeV$^2$; one sample had high positron $p_T$ and the other sample
had high $y$. The variable $y$ is the fraction of the lepton energy transferred to the proton in 
its rest frame and is a measure of the effective longitudinal momentum. Using the two samples, 
full angular coverage of the detector was achieved. The kinematic variables of the positron 
were reconstructed using 
the DA method. The hadronic final state four-vector
was calculated from the EFOs reconstructed as above and its momentum components balanced with
that of the scattered positron. The CAL-EFOs were corrected for energy loss as a function 
of the cluster energy in several angular regions (reflecting the detector 
geometry). The difference between $p_T$ and $y$ for the hadronic system and scattered 
positron was minimised 
and correction factors obtained separately for data and {\sc Ariadne}\cite{ariadne} and 
{\sc Herwig}\cite{herwig} MC simulations as shown in figure~\ref{fig:function}. 
It can be seen that the data and MC show similar trends but differ in detail, justifying 
the need to perform the fits and apply the corrections separately for data and MC.

\begin{figure}[thp]
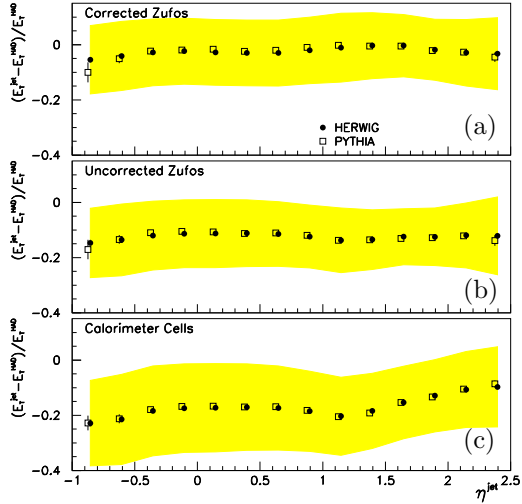

\caption{Fractional difference between hadron-level jet $E_T$ and that reconstructed with 
(a) corrected EFOs, (b) uncorrected EFOs and (c) calorimeter cells as a function of the 
transverse energy. The shaded band shows the width of the distribution. \label{fig:had}}
\end{figure}

To test the validity of the procedure, the correction functions were applied to an
independent photoproduction MC sample, where the scattered positron is not detected in the CAL.
Jet quantities were reconstructed using both EFOs with and without correction and the
transverse energy, $E_T^{\rm jet}$, compared to the hadron-level, $E_T^{\rm HAD}$
as shown in figure~\ref{fig:had}.

\begin{picture}(0,0)(100,100) 
\put(330,-70){(c)}
\put(330,-13){(b)}
\put(330,49){(a)}
\end{picture}
\centerline{\epsfxsize=6.8cm\epsfbox{res.pt3.epsi}}
\vspace{-0.85cm}

Using calorimeter cells as shown in figure~\ref{fig:had}c, the deviation from the true 
value is $20\%$ which is reduced to $10-15\%$ when using EFOs due to the use of tracking 
information as shown in figure~\ref{fig:had}b. After correction, as in fig.~\ref{fig:had}a, 
the transverse energies are significantly closer to the true values, demonstrating that the 
energy correction helps to reproduce the true quantities when applied to an independent MC
sample. To determine the jet energy scale uncertainty, the difference between data and MC 
after the application of these corrections was considered; this is discussed in 
section~\ref{sec:result}.

\subsection{Method 2}

\begin{figure}[th]
\centerline{\epsfxsize=7cm\epsfbox{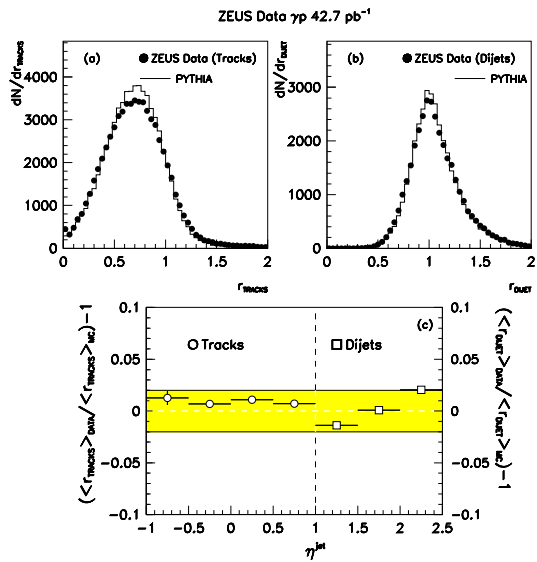}}
\caption{Comparison of (a) $r_{\rm TRACKS}$ and (b) $r_{\rm DIJET}$ for data (points) 
and MC simulation (histogram) and (c) the difference between data and MC as a function 
of pseudorapidity.\label{fig:tracks}}
\end{figure}

In the second method jets are reconstructed using calorimeter cells and a correction for 
energy loss is derived from MC simulation\cite{desy01219}. The reconstructed jet energies are 
corrected on average to the value of the jets from hadrons as a function of transverse 
energy and in regions of pseudorapidity. The correction factors are applied 
to both data and MC events. After this procedure the calorimetric jets in the data and 
MC simulation are compared by utilising tracking information in a cone around the jets. 
The ratio $r_{\rm TRACKS}$ of the jet transverse energy, $E_T^{\rm jet}$, and transverse 
energy of tracks in a cone around the jet axis is shown for data and 
MC in figure~\ref{fig:tracks}a. This quantity can only be calculated within a certain 
angular region corresponding to good acceptance for the central tracking chamber. For a 
jet outside this region, the ratio, $r_{\rm DIJET}$, of its transverse energy to 
that of a central jet was calculated; this is shown in figure~\ref{fig:tracks}b. The mean 
value in data and MC for these ratios was found in different regions of pseudorapidity of 
the jet, $\eta^{\rm jet}$; the difference between data and MC is shown in 
figure~\ref{fig:tracks}c. The MC agrees with the data to within $\pm 2\%$; this deviation 
is then used as a further correction. This procedure relies on the accurate simulation of 
charged tracks and so is less reliant on simulating the energy loss of neutral particles in 
inactive material.
This jet-correction procedure was applied to an independent sample of neutral current DIS 
events. As for method 1, the ratio of the transverse momentum of the positron and hadronic 
jet was calculated and the difference between data and MC determined.

\subsection{Jet energy scale uncertainty}
\label{sec:result}

The jet energy scale uncertainty is shown in figure~\ref{fig:data-mc} as a function of 
pseudorapidity (for method 1) and transverse energy (for method 2). The difference 
between data and MC is within $\pm 1\%$.

\begin{picture}(0,0)(100,100) 
\put(145,75){(a)}
\put(280,75){(b)}
\end{picture}
\centerline{{\epsfxsize=5.cm\epsfbox{ptdacellzufo.epsi}}{\epsfxsize=5.5cm\epsfbox{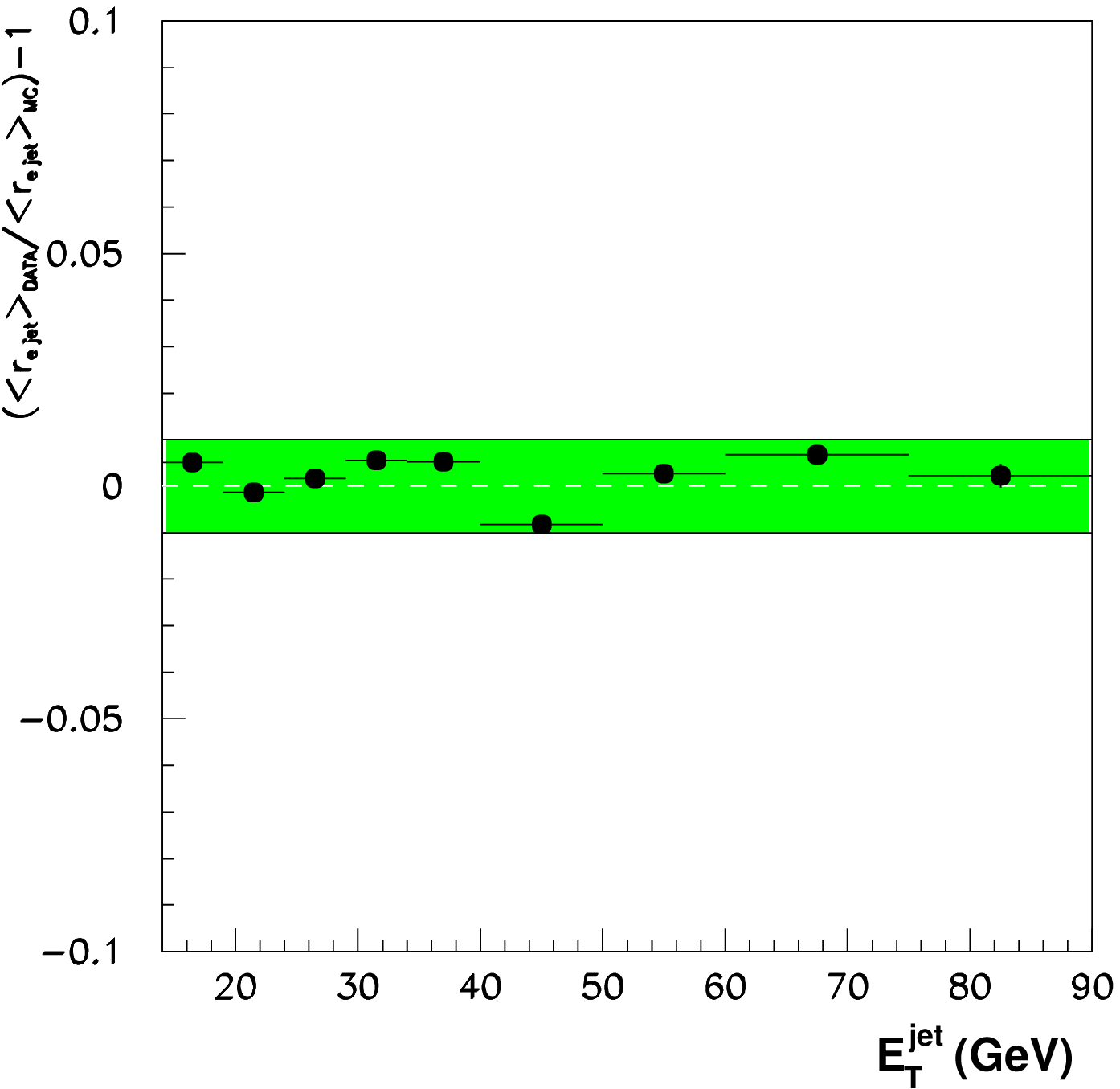}}}
\vspace{-0.5cm}
\begin{figure}[th]
\caption{Jet energy scale uncertainty as a function of (a) $\eta^{\rm jet}$ and 
(b) $E_T^{\rm jet}$. \label{fig:data-mc}}
\end{figure}

\section{Conclusions}

Two independent methods have been developed for correcting jet energies for energy loss 
in inactive material in the detector. Both methods give an improved reconstruction of 
the hadronic final state and understanding of the jet energy scale. The uncertainty of the 
jet energy scale for \mbox{$E_T^{\rm jet} > $ 10 GeV} is $\pm 1\%$. This leads to 
uncertainties in measured cross sections of $\sim \pm 5\%$, significantly smaller than 
current theoretical uncertainties.

\end{document}